\documentclass[%
 reprint,
 amsmath,amssymb,
 aps,
 showkeys
]{revtex4-2}

\usepackage{graphicx}
\usepackage{dcolumn}
\usepackage{bm}

\usepackage{booktabs}

\usepackage{gensymb}
\usepackage{siunitx}

\begin{document}

\preprint{}

\title{The Low Energy Module (LEM): development of a CubeSat spectrometer for sub-MeV particles and Gamma Ray Burst detection}

\author{Riccardo Nicolaidis}
\email{riccardo.nicolaidis@unitn.it}
\author{Francesco Nozzoli}
\author{Roberto Iuppa}
\author{Francesco Maria Follega}
\author{Veronica Vilona}
 \affiliation{Physics Department, University of Trento, 38123 Trento, Italy. \\
 INFN-Trento Institute of Fundamental Physics and Application, 38123 Trento, Italy.}

\author{Giancarlo Pepponi}
\author{Pierluigi Bellutti}
\author{Evgeny Demenev}

\affiliation{Fondazione Bruno Kessler, 38123 Trento, Italy}

\date{\today}

\begin{abstract}
Accurate flux measurement of low energy charged particles, trapped in the magnetosphere, is necessary for Space Weather characterization and to study the coupling between the lithosphere and magnetosphere, allowing the investigation of the correlations between seismic events and particle precipitation from Van Allen Belts. In this work, the project of a CubeSat space spectrometer, the Low Energy Module (LEM), is shown. The detector will be able to perform an event-based measurement of energy, arrival direction, and composition of low-energy charged particles down to 0.1 MeV. Moreover, thanks to a CdZnTe mini-calorimeter, the LEM spectrometer also allows photon detection in the sub-MeV range, joining the quest for the investigation of the nature of Gamma Ray Bursts. The particle identification of the LEM relies on the $\Delta E - E$ technique performed by thin silicon detectors. This multipurpose spectrometer will fit within a 10x10x10 $\text{cm}^3$ CubeSat frame and it will be constructed as a joining project between the University of Trento, FBK, and INFN-TIFPA. To fulfil the size and mass requirements an innovative approach, based on active particle collimation, was designed for the LEM, this avoids heavy/bulky passive collimators of previous space detectors. In this paper, we will present the LEM geometry, its detection concept, and the results from the developed GEANT4 simulation.
\end{abstract}

\maketitle

\section{Introduction}

The Low Energy Module (LEM) will be a compact spectrometer able to perform an event-based measurement of energy, direction and composition of low energy charged particles, in particular down to 0.1 MeV for electrons.
The physics goal of this detector is the monitoring of the magnetosphere and ionosphere environment. It is known that the measurements of the fluxes of low energetic particles could allow the characterisation of the coupling between the lithosphere, atmosphere, ionosphere, and the magnetosphere. In particular, earthquakes are dynamical processes caused by a continuous and slow strain accumulation. From studies on fault rupture mechanics, seismic wave propagation, and geophysical parameters measured in the ionosphere and the low magnetosphere, some anomalies correlated with catastrophic events were discovered. Moreover, also a statistical evidence of temporal correlation between particle precipitations from Van Allen Belts and strong seismic events has been pointed out \cite{battiston2013first}. These observations, motivate the interest in further, detailed, measurements of electron fluxes in the energy window $0.1 - 7 \ \text{MeV}$ that could be a promising channel to identify possible seismic precursors. 

Another interesting case study for the LEM instrument is the application to space weather. Severe space weather storms could cause power outages and telecommunication alterations. 
For this reason, the construction of new instruments to monitor and (possibly) to predict the effects of solar activity on Earth is crucial. 

The LEM will be a particle telescope performing an event-based measurement of energy,
direction and composition of low energy charged particles, in particular electrons down
to 0.1 MeV. This capability is not possible with the existing detectors that fails the possibility of an event-based PID or fails the possibility of monitoring the particle flux from
different directions at the same time, or are not able to measure directions of low energy particles
because of the multiple scattering occurred in the first layer of a particle-tracking configuration.


\section{The current landscape of space-based particle detectors}

There is an extensive literature about particle detectors in space using silicon technologies. The instruments under examination, are the following: The Instrument for the Detection of Particles (IDP) on the DEMETER microsatellite \cite{parrot2015demeter, sauvaud2006high, dubourg2006demeter}, The High Energy Particle Package (HEPP-H and HEPP-L) on CSES \cite{xu2010design, wu2013design, HEPPHL, hephonly}, the Mars Energetic Particle Analyser (Mars-EPA) on the Tianwen-1 mission \cite{li2021design, tang2020calibration, zou2021scientific}, and the Radiation Assessment Detector on the Curiosity rover \cite{hassler2012radiation, zeitlin2016calibration, ehresmann2014charged, ehresmann2016charged, guo2015msl}.

\begin{table*}[ht]
\caption{Summary of some features of the detectors studied in this section. The references from which I extracted the information are quoted within the text.}
\centering
    \begin{tabular}{llllllll}
    \toprule
        Instrument & Size    & Directions & Angular    & Energy  & PID & Detector\\
                   & Weight  &            & Resolution & range   &   & Elements \\
        \midrule
        IDP DEMETER   & 525 g                 & 1                    & FOV 32 deg.   & e: [0.07, 0.8] MeV     & No  & Silicon Diode\\
        \midrule
        RAD Curiosity & $\sim$10x10x10 cm$^3$ & Complex              & FOV 36.7 deg. & e: [0.1, 20] MeV       & Yes & PIPS (3 segments)\\
                      &                       & segmentation         &               & p: [5, 200] MeV        &     & CsI(Tl)\\
                      &                       &                      &               & $\alpha$: [5, 200] MeV &     & Plast. Scint.\\                      
                      &                       &                      &               & l.Z: [10, 300] MeV     &     &  \\        
        \midrule
        HEPP-L        & Large                 &  - 5 Narrow          & - FOV 6.5 deg.& e: [0.1, 3] MeV & Yes  & Si det. (2 layers)\\
                      & Collimators           &  - 4 Wide            & - FOV 15 deg. & p: [2, 20] MeV  &      & Plast. Scint.\\
        \midrule
        Mars-EPA     &  270x180x148 cm$^3$   & 1                    & FOV 60 deg.   & e: [0.1, 12] MeV   & Yes & PIPS (2 layers)\\
                      &                       &                      &               & p: [2, 100] MeV    &     &  CsI(Tl)\\
                      &                       &                      &               & $\alpha$: [25, 400] MeV    & &    \\
                      &                       &                      &               & l.Z: [25, 400] MeV    &    & \\
        \bottomrule
    \end{tabular}
    \label{tab:review_detector}
\end{table*}

Some of the most important features of these instruments are listed in table \ref{tab:review_detector}. Even though all of these experiments have different scientific purposes and goals, their detection concept and scheme are very similar, allowing a comparison between their structure, size, components, and performances. In table \ref{tab:review_detector}, the reader can see a summary of the features characterising the previously mentioned experiments. By comparing the six detectors studied, we can conclude that the larger is the number of layers inserted into the design, the better are the performances in detecting energetic particles. Furthermore, by adding additional layers at the bottom of the instrument, as we can see in the RAD, on the Curiosity mission, or in the Mars-EPA, the maximum energy that can be detected with particle identification is increased. On the other hand, to minimise the low energy threshold, one has to minimise the thickness of the $\Delta E$ layer.
As an example, the Mars-EPA, can detect electrons in the energy range of $0.1-2 \ \text{MeV}$ by adopting a $\Delta E$ layer made of Passivated Implanted Planar Silicon (PIPS) detectors with a thickness of $15\ \mu\text{m}$. Finally, the use of an inorganic scintillator as a calorimeter could be problematic. In particular, many scintillator crystals like Sodium-Iodide or Caesium-Iodide, are very fragile and hygroscopic. These aspects will unavoidably result in the introduction of mechanical supports or metallic wrapping providing additional dead layers in which particles could deposit part of their energy. 

Furthermore , it is required that the LEM is compact (within 10x10x10cm$^3$) and that it can monitor the particle flux in a large field of view from different directions at the same time. 
These capabilities are not simultaneously fulfilled by the past detectors. Therefore, a different and innovative design is required for the LEM.

\section{The LEM concept: the active collimation technique}

The idea that allows the reduction of weight and size of the LEM detector relies upon the active collimation technique. More precisely, a drilled plastic scintillator is acting as a veto.
Only particles with directions aligned with one of the 16 channels are detected by one of the 16 silicon sensor pairs. Thus, direction information is obtained.
Particles with wrong direction are stopped in the Aluminium shield or will release a signal in the drilled plastic scintillator veto.
This technique is an alternative to the tracking one affected by the multiple scattering problem.
On the other hand, the low density of the plastic scintillator veto avoids the significant weight required by a totally passive metallic collimator. However, the price to pay is a relatively high veto rate. This high veto rate will unavoidably result in a enhancement of the dead-time of the detector. For this reason, a small drilled aluminium shield is still necessary to suppress very low-energy particles. 

\begin{figure*}[ht]
    \centering
    \includegraphics[width = 0.9 \linewidth]{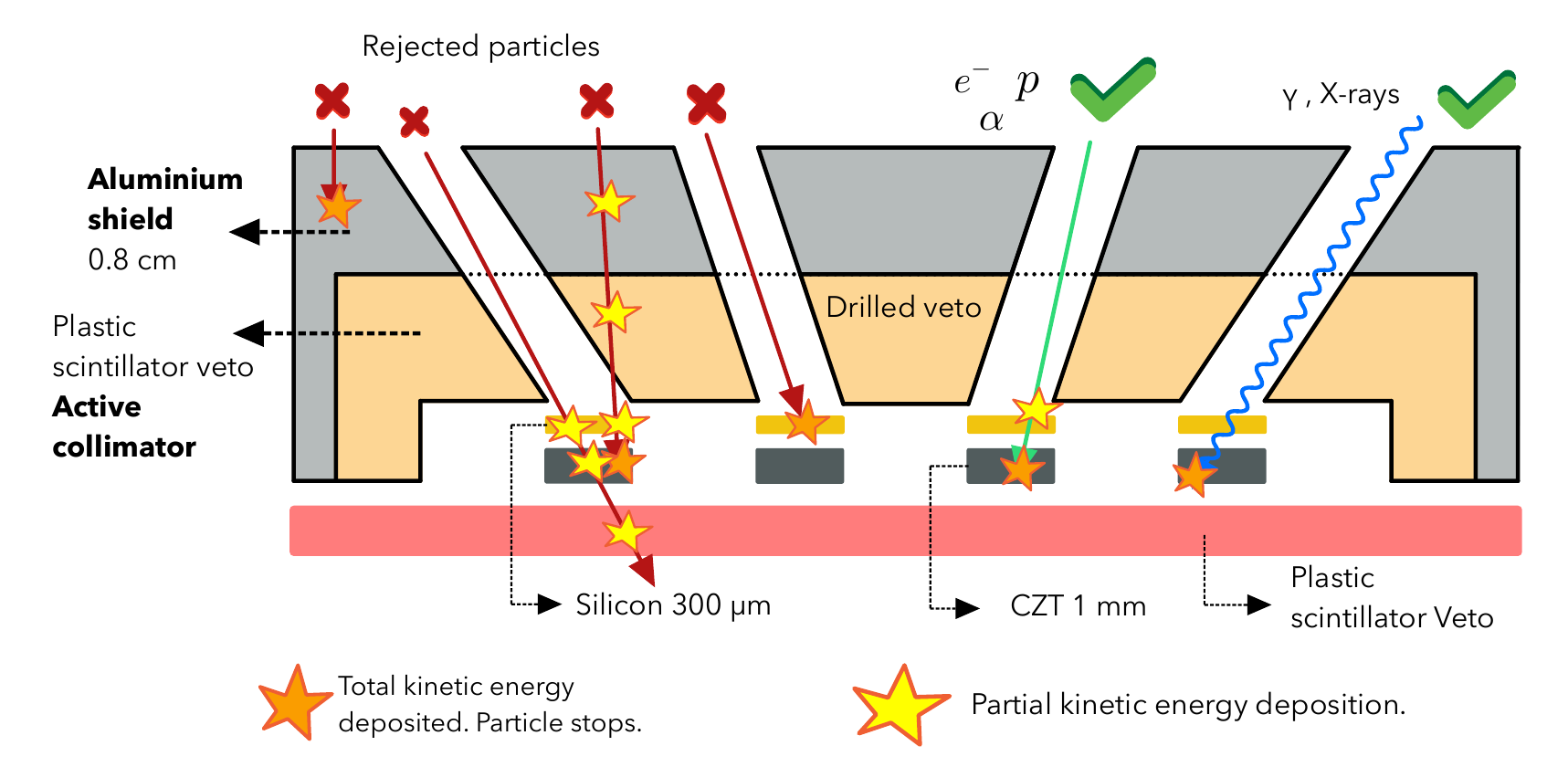}
    \caption{Detection concept embedded within the detector's geometry. The legend allow to distinguish a partial energy deposition from a total kinetic energy deposition.}
    \label{fig:lem_geometry_dee}
\end{figure*}

In figure \ref{fig:lem_geometry_dee}, the detection concept and a schematised cross section of the instrument are shown. From the top, we can see the drilled aluminium mask suppressing the flux of very low energetic particles. Below the aluminium shield, the active anti-coincidence is obtained by using a drilled plastic scintillator (polyvinyl toluene). The aluminium drilled mask and the drilled Anti-Coincidence Detector (ACD) define the so-called active collimator.

For a LEM operating in Low Earth Orbit (LEO), an Aluminium thickness larger than $0.5 \ \text{cm}$ is necessary to reduce the veto rate from several MHz to the affordable rate of $\sim \text{kHz}$. Below the active collimation system, we place the 16 independent $\Delta E - E$ modules. These $\Delta E - E$ modules will measure the angular flux of particles crossing the veto channels ($\oslash 1$cm $\times$ 1.3 cm), encoding one specific solid angle in the sky with a resolution of $6\degree-7\degree$. The size of commercially available PIPS detectors manufactured by Ametek (50mm$^2$ - $\oslash$ 8mm each) have been considered to define a realistic geometry in the detector simulation. The $\Delta E$ detector consists of a $100 \ \mu \text{m}$ thick PIPS detector while, the $E$ detector is a CdZnTe detector with $1 \ \text{mm}$ thickness. 
These two $\Delta E$ - $E$ layers, allow a good particle identification in the energy ranges: about $0.1 - 10 \ \text{MeV}$ for electrons, $3 - 30 \ \text{MeV}$ for protons, and $10 - 100 \ \text{MeV}$ for alpha particles. A bottom plastic scintillator (ACD) is added at the very end of the LEM to ensure the energy release is confined within the above layers. In particular, Particle Identification (PID) is not possible both for energetic particles crossing the ACD as well as for slow particles stopped in the front PIPS.  
Events with undefined direction are rejected thanks to a signal released in the active veto/collimator.
Finally, events which are fully contained within the LEM, are selected. In this very last case, the direction is well defined and it is also possible to perform an accurate PID. 
Thanks to the high density and high averaged atomic number of CZT \cite{takahashi2001recent, sordo2009progress}, the LEM can identify low energy $\gamma$-rays converting in the CZT mini-calorimeter and using all the surrounding low-Z sensors as anticoincidence. 
The ability of observing energetic photons will allow to use this compact particle spectrometer also as a Gamma-Ray Burst (GRB) monitor \cite{piran2005physics}.

\section{Performance characterisation with GEANT 4 simulation}

The detection concept adopted in the LEM is a consolidated technique denominated $\Delta E- E$ \cite{scarduelli2020method, evensen1997thin, carboni2012particle}. Basically, a $\Delta E - E$ particle spectrometer is composed of a thin detection layer and a thicker one behind. When a particle impinges on the spectrometer, if the kinetic energy is enough, the particle could cross the first layer, releasing a part of its kinetic energy $\Delta E$. Then, the residual kinetic energy $E$ could be entirely deposited within a second, thicker, layer. This experimental layout allows particle identification by measuring the energy deposited in the thinner layer, the $\Delta E$ energy, and the energy deposited in the thick sensor, the $E$ energy.
If a sub-MIP particle passes through a thin detector layer, the energy deposited $\Delta E$ will be velocity dependent:
\begin{equation}
    \Delta E \approx \frac{Z^2}{\beta^2}
    \label{eq:approxStoppingPower}
\end{equation}
where $Z$ is the projectile's charge and $\beta$ is its velocity in natural units. On the other hand, also the residual kinetic energy, $E$, of a particle stopping in a subsequent thick detector is velocity dependent:
\begin{equation}
    E = m c^2 (\gamma - 1) \approx \frac{1}{2}m (\beta c)^2
\end{equation}
where the non-relativistic approximation holds for sub-MIP particles. 
Therefore, in a $\Delta E-E$ spectrometer, a useful PID classifier could be defined in the following way:
\begin{equation}
\begin{split}
    \text{PID}_{\text{classifier}} & = \log_{10} \left[ \frac{\Delta E}{1 \ \text{MeV}} \frac{E}{1 \ \text{MeV}} \right] \approx \\
    & \approx \text{constant} + \log_{10} Z^2 \left( \frac{m c^2}{1 \ \text{MeV}}\right) \\
\end{split}
    \label{eq:PID_definition}
\end{equation}

\begin{figure*}[ht]
    \centering
    \includegraphics[width = 0.85 \linewidth]{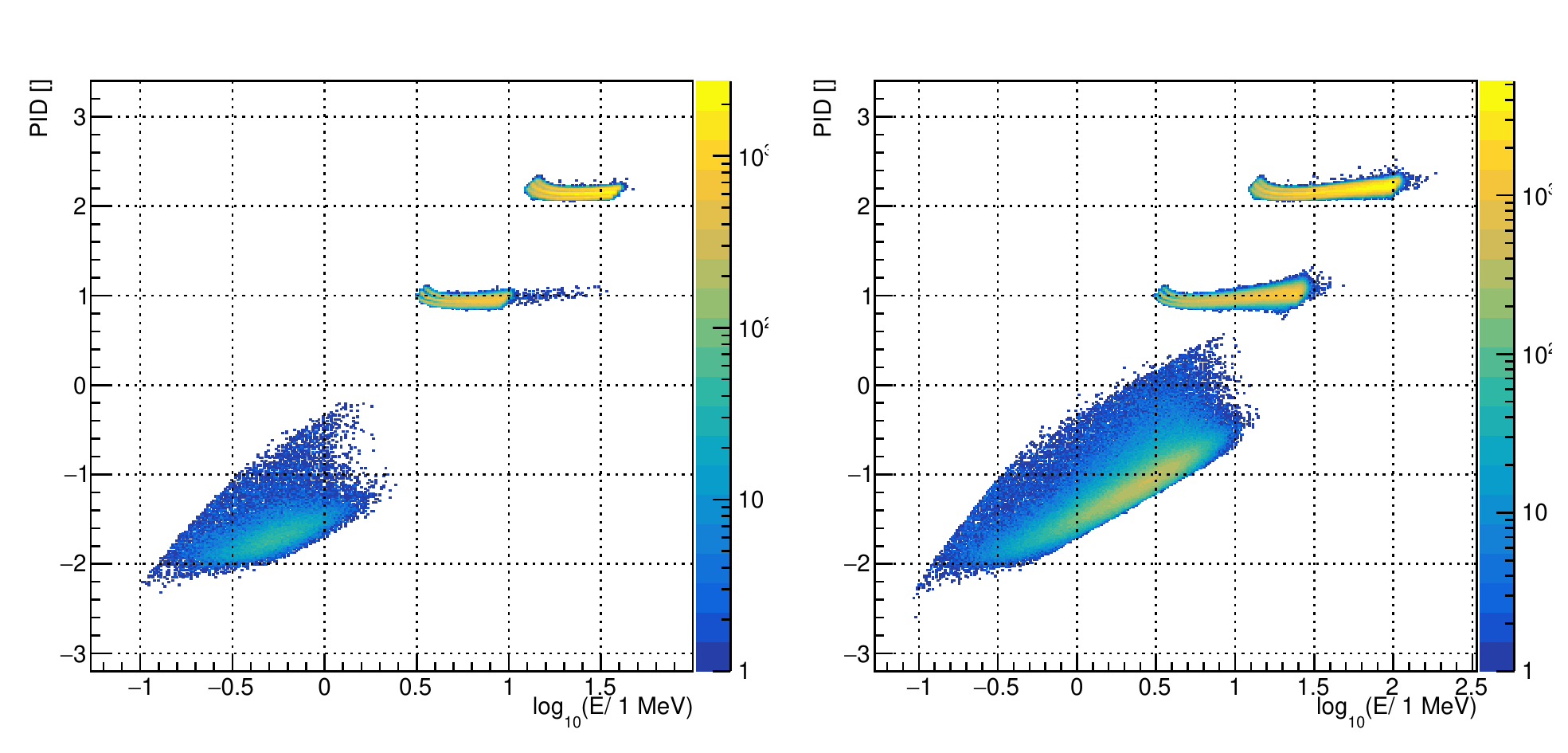}
    \caption{PID classifier vs kinetic energy. Left: $100 \ \mu\text{m} - 500 \ \mu \text{m}$ PIPS detectors. Right: $100 \ \mu \text{m}$ PIPS and $ 1 \ \text{mm}$ CZT detectors. The three different clusters in each plot represent (from the top to the bottom) : alpha particles, protons, electrons.}
    \label{fig:CZTSiComparisonPID}
\end{figure*}

\begin{figure}[h]
    \centering
    \includegraphics[width = 1. \linewidth]{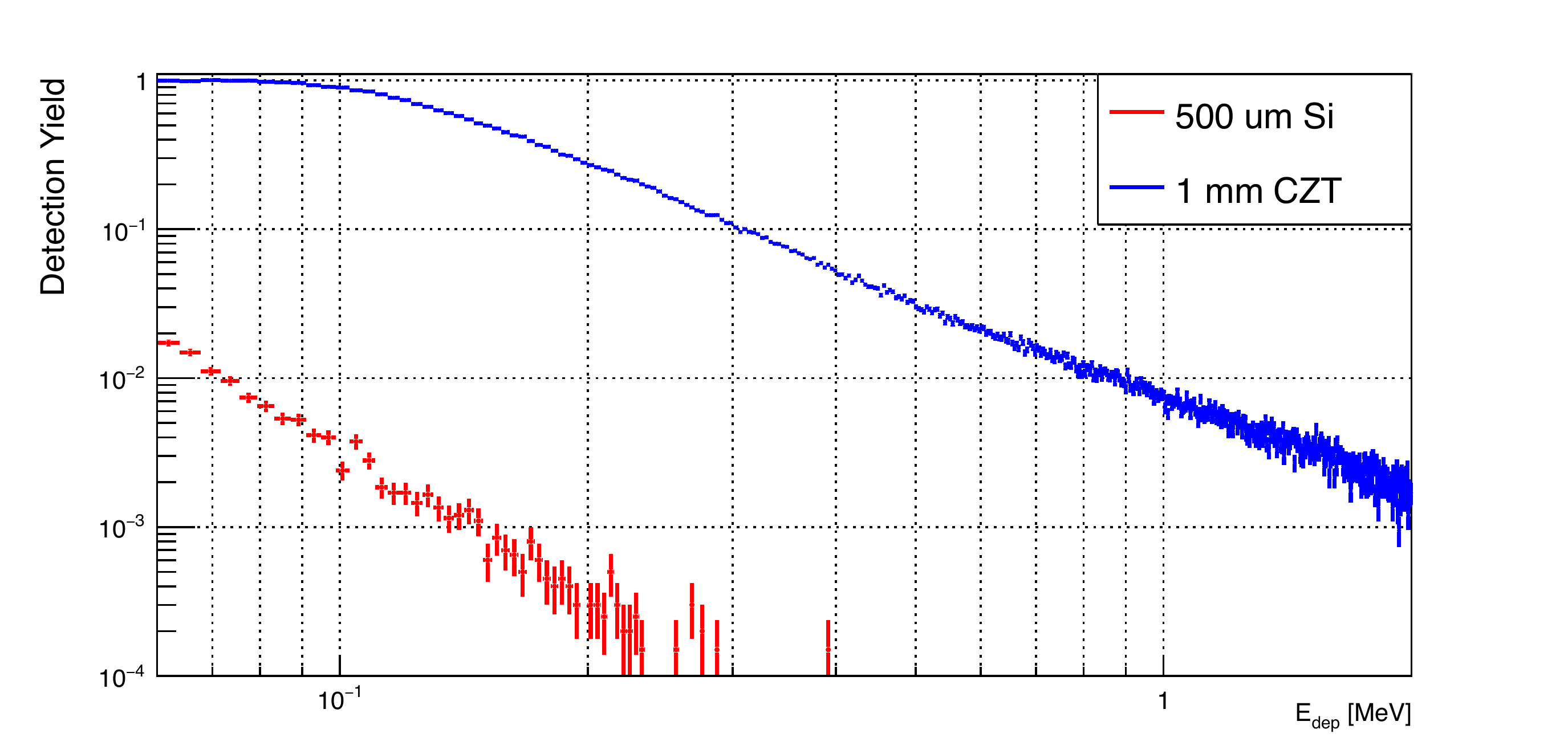}
    \caption{Improvement of detection yield for $\gamma$-rays thanks to CZT sensors. The legend refers to the material and the thickness considered for the $E$ detector.}
    \label{fig:CZTSiComparisonGamma}
\end{figure}

This $\text{PID}_{\text{classifier}}$ allows to remove the main dependence on the particle's velocity for sub-MIP particles, therefore it is mainly dependent only from the particle's nature. For the LEM detector, this approximation is very good for protons and other nuclei but for electrons in the LEM kinetic energy range the sub-MIP approximation fails and PID classifier for electrons will grow roughly like $\log_{10} \frac{E}{1 \ \text{MeV}}$. However a good identification of electrons from protons based on this classifier is still achieved thanks to the fact that proton mass is 2000 times larger than electron mass. In figure \ref{fig:CZTSiComparisonPID}, the results from a GEANT4 \cite{agostinelli2003geant4} simulation are shown. 
In particular, the PID vs Energy identification capability for the case of a mini-calorimeter made of $500 \ \mu \text{m}$ Silicon (left plot) is compared with the case of 
a mini-calorimeter composed by 1 mm thick CZT.

Finally, to quantify the advantage of the use of CZT sensors in the LEM as a monitor for GRBs, a comparison of the relative photon detection efficiency for the two mini-calorimeter configurations is shown in Figure \ref{fig:CZTSiComparisonGamma}. 

\section{Conclusion and Outlooks}

In this work we described the Low Energy Module (LEM): a compact particle spectrometer, suitable for a CubeSat , for measurement of the differential flux of low-energy particles in the lower magnetosphere. 
Here, it is worth summarising the structure of the LEM. To avoid a bulky and heavy detector,
we design an active collimator based on a thin Aluminium shield followed by an anti-coincidence detector. The drilled aluminium shield protects the drilled ACD, made of a plastic scintillator, from the large flux of very low energetic electrons in LEO. The holes in the aluminium and in the ACD are used to select a known direction of the particles with an angular resolution of $6\degree-7\degree$. The LEM Field of View is $60\degree \times 60\degree$ monitoring 16 directions in the sky at the same time. 
The particle identification relies in a series of 16 $\Delta E - E$ modules based on PIPS and CdZnTe detectors placed below each collimator channel. An additional layer of plastic scintillator at the bottom is added as a veto to identify non-contained, particles. 
%
%
Test on sensor prototypes are ongoing at INFN-TIFPA laboratory.

\bibliography{apssamp}

\end{document}